\documentclass[fleqn,twoside]{article}%
\usepackage{amsmath}
\usepackage{amsfonts}
\usepackage{amssymb}
\usepackage{graphicx}
\usepackage{caption}
\usepackage{cite}

\def\be{\begin{equation}}
\def\ee{\end{equation}}
\def\bi{\bibitem}
\begin{document}
\title{Inflation-a comparative study amongst different modified gravity theories.}
\author{Dalia Saha$^1$ and Abhik Kumar Sanyal$^2$}
\maketitle
\noindent
\begin{center}
\noindent
$^{1, 2}$ Dept. of Physics, Jangipur College, Murshidabad, West Bengal, India - 742213\\
\end{center}
\footnotetext[1]{
\noindent Electronic address:\\
$^1$daliasahamandal1983@gmail.com\\
$^2$sanyal\_ ak@yahoo.com\\}
\begin{abstract}
In the recent years, a host of modified gravity models have been proposed as alternatives to the dark energy. A quantum theory of gravity also requires to modify `General Theory of Relativity'. In the present article, we consider five different modified theories of gravity, and compare inflationary parameters with recent data sets released by two Planck collaboration teams. Our analysis reveals that the scalar-tensor theory of gravity is the best alternative.
\end{abstract}
\section{Introduction:-}

After some initial debate, cosmologists have unanimously and unambiguously come to a very weird conclusion that the universe is currently accelerating. Weird, since gravity is attractive, a fifth force (quintessence) must be responsible for such a phenomena. General Theory of Relativity (GTR) described by the equation

\be\label{I1} G_{\mu\nu} = R_{\mu\nu} - {1\over 2}g_{\mu\nu}R = \kappa T_{\mu\nu},\ee
where left hand side is the Einstein tensor which describes the curvature of space time, and the right hand side is the energy-momentum tensor of baryonic matter and non-baryonic dark matter, $\kappa = 8\pi G$, $G$ being the Newton's gravitational constant; can not address such phenomena. The reason being: the equation of state parameter is $\omega = {p\over\rho} \ge 0$ (where $p$ and $\rho$ are the thermodynamic pressure and matter density respectively), while accelerated expansion of the universe requires a negative pressure, so that the equation of state parameter is $\omega_e < -{1\over 3}$, where, the subscript `${e}$' stands for `effective'. To be  precise, current data suggests $\omega_e < -{2\over 3}$. Therefore, GTR somehow, has to be modified. Cosmological constant ($\Lambda$), for which $\omega_\Lambda = {p_\Lambda\over \rho_\Lambda} = -1$, can resolve the issue singlehandedly, but then, what is a cosmological constant? A physical interpretation of it comes from high energy physics, in which one can compute a `constant' available in the nature, as the sum of vacuum energy densities of all types of matter existing in the universe. Unfortunately, the constant required for current acceleration of the universe is $120$ order of magnitude smaller than the sum of vacuum energy densities. Thus, $\Lambda \mathrm{CDM}$ (cold dark matter) model was replaced initially by a quintessence field, which is essentially a scalar field, for which $\omega_{e} = {p_e\over \rho_e} = {{1\over 2}\dot\phi^2 - V(\phi)\over {1\over 2}\dot\phi^2 + V(\phi)}$. Clearly, quintessence model does not admit the value of the equation of state parameter to go beyond the phantom divide line, $\omega = -1$, since if $\dot\phi^2 \ll V(\phi)$, then $\omega_e$ takes the limiting value, $\omega_e = -1$. However, crossing of the phantom divide line is not excluded by observations. Therefore, different exotic models (K-essence, Tachyon, holographic model etc.) were proposed. These are all dark energy models, since such fields interact none other than gravity itself. These fields essentially modifies the energy-momentum tensor ($T_{\mu\nu}$), that is the right hand side of Einstein's equation of GTR. However, since all attempts to detect dark energy has failed\footnote{There is a very recent indication of direct detection of dark energy in XENON1T, that we shall discuss in brief in the conclusion.}, so the cosmologist started modifying the left hand side of Einstein's equation, namely the curvature part. Einstein's equation of GTR \eqref{I1} may be found under the variation of the so-called Einstein-Hilbert action,

\be\label{12} A = \int\left[ {R\over 16\pi G}\right]\sqrt{-g} d^4x + S_m,\ee
where, $R$ is the Ricci scalar, $-g$ is the determinant of the metric, and $S_m$ is the matter action. In order to modify the left hand side of Einstein's equation \eqref{I1}, it is required to replace Ricci scalar $(R)$ by a generalized curvature scalar. A host of such models $F(R),~F(\mathcal{G}),~F(T),~F(Q)$ etc., where, $\mathcal{G} = R^2 + 4R_{\mu\nu}R^{\mu\nu} + R_{\mu\nu\delta\gamma}R^{\mu\nu\delta\gamma}$ is the Gauss-Bonnet term, $T$ is the torsion term, $Q$ is the non-metricity scalar, have so far been proposed. All these models can address late-time cosmic acceleration followed by an early decelerating phase. On the other hand, construction of a quantum theory of gravity also requires to modify GTR, by incorporating higher order curvature invariant terms in the gravitational action \cite{Stelle}. It is therefore suggestive to check if these modified theories can explain early stage of cosmic evolution, an inflationary phase, in particular.\\

Standard model of cosmology, the so-called `Friedmann-Lemaitre-Robertson-Walker (FLRW) model' predicts that the universe initiated from a big-bang, represented by a hot thick soup of plasma. The evidence of extremely hot big-bang has been experimentally verified through the detection of CMBR (Cosmic Microwave Background Radiation). However, causally disconnected regions appear to be isotropic up to $10^{-5}$ order of magnitude, called the horizon problem, which is not explained by FLRW model. Further, FLRW model does not explain flatness problem (the fact that the universe is almost flat at present, and a slightest deviation, would have collapsed it very early, or would have enormously expanded it, desisting to form structures). Finally FLRW model does not also account for the structure (stars, galaxies, cluster of galaxies etc.) formation. All these issues may be addressed if there had been a stage of inflation (exponential or power law expansion of the scale factor $a(t)$) in the very early stage of cosmological evolution \cite{inflation1, inflation2, inflation3, inflation4, inflation5}. Although, there exists some models which appear to explain these issues \cite{alternative1, alternative2, alternative3}, inflation is prevalent, mainstream choice, and is considered to be a scenario, rather than a model. In this connection, it is suggestive to check, if proposed modified theories of gravity can accommodate inflation as well. Inflation is essentially a quantum phenomena, which was initiated sometime between ($10^{-42}~ \mathrm{to}~ 10^{-26}~s$), after gravitational sector transits to the classical domain. To be more specific, it is a quantum theory of perturbations on top of a classical background, which means the energy scale of the background must be much below Planck scales. There are also recent evidence from the string theory swampland that the energy scale must be rather low for inflation. Despite the fact that inflation is a quantum phenomena, most of the important physics may be extracted from the classical action itself, provided the quantum theory admits a viable semiclassical approximation. We have shown earlier that the models under consideration, admit viable quantum dynamics and are classically allowed, since the semiclassical wave-functions oscillate about classical inflationary solutions.\\

In view of the above discussions, in the following section of the present article, we consider five different well versed modified theories of gravity to study inflation. In particular, we inspect how far these models fit with the currently released inflationary parameters \cite{Planck1, Planck2}, namely the tensor to scalar ratio $r = 16\epsilon < 0.06$, where $\epsilon$ is the first slow roll parameter, and the scalar tilt, or more conventionally the spectral index of scalar perturbation $0.096 < n_s < 0.097$, and the number of e-folding remains preferably within the range $40 < \mathrm{N} < 70$, required to solve the horizon and the flatness problems. First four of these models are higher order theories, while one appearing at the end, is a non-minimally coupled scalar-tensor theory of gravity. The comparative study that we are going to perform, will render a selection rule to consider a particular modified gravitational action. In section 3 we conclude.

\section{Inflation in different modified theories of gravity:}

So far, all attempts to cast a viable (although unsuccessful) quantum theory of gravity addressed higher order scalar curvature square terms ($R^2,~R_{\mu\nu}R^{\mu\nu}$) in the action. Further, Gauss–Bonnet term $\mathcal{G} = (R^2 + 4R_{\mu\nu}R^{\mu\nu} + R_{\mu\nu\delta\gamma}R^{\mu\nu\delta\gamma})$, appears quite naturally as the leading order of the inverse string tension $\alpha'$ expansion of heterotic superstring theory \cite{het1, het2, het3, het4}. But, Gauss-Bonnet term is topologically invariant in $4$-dimension, which means, it is a total derivative term, and therefore does not contribute to the field equations. However, when coupled to a scalar field (dilaton), it contributes. In this context, it is noteworthy that the low energy limit of the string theory gives rise to the dilatonic scalar field, which is also found to be coupled with various curvature invariant terms \cite{dil1, dil2}. Therefore, the leading quadratic correction gives rise to Gauss–Bonnet term with a dilatonic coupling \cite{dil3}. It is important to mention that the dilatonic coupled Gauss–Bonnet term plays a vital role at the late-stage of cosmic evolution (pressure-less dust era), exhibiting accelerated expansion after a long Friedmann-like decelerating phase \cite{acc1, acc2}. The higher order theories under consideration therefore should contain these terms in different combinations. \\

We shall work in the following homogeneous and isotropic Robertson-Walker metric, viz.,

\be \label{RW} ds^2 = -dt^2+a^2(t)\Big[\frac{dr^2}{1-kr^2}+r^2(d\theta^2+sin^2\theta\; d\phi^2)\Big],\ee
where, $a(t)$ is the scale factor. The Ricci scalar and the Gauss-Bonnet term for the above space-time \eqref{RW} are given by,

\be \label{R}\begin{split} {R}&=6\Big(\frac{\ddot{a}}{a}+\frac{\dot{a}^{2}}{a^2}+\frac{k}{a^{2}}\Big),~~~~~
\mathcal{G} = {24}{\ddot a\over a^3}\left({\dot a^2} + k\right),\end{split}\ee
which we shall require to cast the field equations.

\subsection{Case-1:}

First, we start with the following generalised action considered earlier in \cite{In1}

\be\label{A1}
A_1=\int d^{4}x \sqrt{-g} \bigg{[}{\alpha(\phi)R}-\Lambda M^2_P+\beta(\phi)R^2+\gamma(\phi)\mathcal{G}-\frac{1}{2}\phi_{,\mu}\phi^{,\mu}-V(\phi)\bigg{]}. \ee
The action contains undetermined coupling parameters $\alpha(\phi),~\beta(\phi),~\mathrm{and}~\gamma(\phi)$ and a cosmological constant term ($\Lambda$) being coupled to the reduced Planck's mass $M^2_P = {1\over 8\pi G}$.

\subsubsection{Field equations and classical solutions:}

The field equations, namely the `$a$' variation i.e. ($^i_i$) equation, the $(^0_0)$ equation and the $\phi$ variation equation for the metric \eqref{RW} are the following,

\be \label{avariation1} \begin{split}&2\alpha\bigg[\frac{2\ddot{a}}{a}+\frac{\dot{a}^2}{a^2}+\frac{k}{a^2}\bigg]+2\alpha'\bigg[\ddot{\phi}+\frac{2\dot{a}\dot{\phi}}{a}\bigg]
+2\alpha''\dot{\phi}^2+12\beta\bigg[\frac{2\ddddot{a}}{a}+\frac{4\dot{a}\dddot{a}}{a^2}+\frac{3\ddot{a}^2}{a^2}-\frac{12\dot{a}^2\ddot{a}}{a^3}
+\frac{3\dot{a}^4}{a^4}-\frac{4k\ddot{a}}{a^3}+\frac{2k\dot{a}^2}{a^4}-\frac{k^2}{a^4}\bigg]\\& \hspace{1.2 cm}+48\beta'\dot{\phi}\bigg[\frac{\dddot{a}}{a}+\frac{2\dot{a}\ddot{a}}{a^2}-\frac{\dot{a}^3}{a^3}-\frac{k\dot{a}}{a^3}\bigg]+(24\beta''\dot{\phi}^2+24\beta'\ddot{\phi})\bigg[\frac{\ddot{a}}{a}+\frac{\dot{a}^2}{a^2}+\frac{k}{a^2}\bigg]
+\frac{16\gamma'\ddot{a}\dot{a}\dot{\phi}}{a^2}+8\gamma'\ddot{\phi}\bigg[\frac{\dot{a}^2}{a^2}+\frac{k}{a^2}\bigg]\\&\hspace{6.1 cm}+8\gamma''\dot{\phi}^2\bigg[\frac{\dot{a}^2}{a^2}+\frac{k}{a^2}\bigg]+\frac{\dot{\phi}^2}{2}-V-\Lambda M^2_P=0.\end{split}\ee

\be \label{001} \begin{split}6\alpha\bigg{(}{\dot a^2\over a^2}+ {k \over a^2}\bigg{)} + 6\alpha'\dot\phi\Big({\dot a\over a}\Big) &+36\beta \bigg{(}\frac{2\dot a\dddot a}{a^2}-\frac{\ddot a^2}{a^2} +\frac{2\dot a^2 \ddot a}{a^3}-\frac{3\dot a^4}{a^4}-{2k\dot a^2\over a^4}+{k^2\over a^4}\bigg{)}\\&+72\beta'\dot\phi\bigg{(}\frac{\dot a\ddot a}{a^2}+\frac{\dot a^3}{a^3}+{k\dot a\over a^3}\bigg{)}+24\gamma'\dot\phi\bigg{(}{\dot a^3\over a^3}+{k\dot a\over a^3}\bigg{)}-\Lambda M^2_P = \bigg{(}\frac{\dot\phi^2}{2}+V \bigg{)}, \end{split} \ee

\be \label{phivariation1} \begin{split}& \ddot\phi +3{\dot a\over a} \dot\phi + V' - 6\alpha'\Big({\ddot a\over a}+{\dot a^2\over a^2} + {k\over a^2}\Big)- 36\beta'\Big({\ddot a^2\over a^2}+2{\dot a^2\ddot a\over a^3}+ \frac{\dot a^4}{a^4}+2{k\ddot a\over a^3}+{2k\dot a^2\over a^4}+{k^2\over a^4}\Big) -24\gamma'\Big({\dot a^2\ddot a\over a^3}+ {k\ddot a\over a^3}\Big)  =0.\end{split}\ee
In the above, and throughout, an over-dot denotes time derivative, while prime denotes derivative with respect to the scalar field. Not all the above components of Einstein's equations are independent, since the $(^0_0)$ equation is the energy constraint equation. Thus it suffices to consider only the two independent components of Einstein's equations, viz. (\ref{001}) and (\ref{phivariation1}), for all practical purposes. A viable gravity theory must admit de-Sitter solution ($a \propto e^{\lambda t}$) in vacuum. As already explored earlier \cite{In1}, the above field equations admit the following de-Sitter solution in the spatially flat space ($k=0$) space-time,

\be\label{aphi1} \begin{split} a = &a_0e^{\lambda t}; ~~~ \phi= \phi_0e^{-\lambda t},~~\mathrm{under~ the~ condition},\\&
\alpha(\phi) = {\alpha_0\over \phi};~~ V(\phi) = \frac{1}{2} \lambda^2 \phi^2-\Lambda M^2_P;~~\mathrm{and}~~ 6\beta(\phi) + \gamma(\phi) = - {1\over 2\lambda^2}\left({\alpha_0\over \phi} +{\phi^2\over 24}\right),\end{split}\ee
where, $a_0$, $\phi_0$, $\alpha_0$ and $\lambda$ are arbitrary constants, while $\beta(\phi)$ and $\gamma(\phi)$ remain arbitrary functions of $\phi$, being related as above, after setting the constant of integration to zero without any loss of generality.

\subsubsection{Inflation under Slow Roll Approximation:}

As mentioned in the introduction, the model under consideration admits a viable (hermitian) quantum dynamics, while the semiclassical wave-function oscillates about the above classical inflationary solution \eqref{aphi1}, and thus it is classically allowed. Hence although Inflation is a quantum mechanical phenomena, most of the important physics are inherent in the classical action. We therefore proceed to study inflation and see how far the inflationary parameters viz. the tensor to scalar ratio $r$ and the spectral index of scalar perturbation $n_s$ fit with currently released data sets $r < 0.06$ and $0.096 < n_s < 0.097$, keeping the number of e-folding within the range $45 < \mathrm{N} < 70$, required to solve the horizon and the flatness problems \cite{Planck1, Planck2}. For a complicated theory such as the present one, it is of-course a very difficult task. However, we followed a unique technique to make things look rather simple, which is described underneath. We express equations \eqref{001} and \eqref{phivariation1} in terms of the Hubble parameter $\mathrm{H}$ in the spatially flat space-time ($k = 0$) as,

\be \label{hir1}\begin{split} 6\alpha \mathrm{H}^2 = \frac{\dot\phi^2}{2} + \left[V+\Lambda M^2_P - \big(6\alpha'\dot\phi\mathrm{H} + 144\beta'\dot\phi\mathrm{H} ^3 + 24\gamma'\dot\phi \mathrm{H}^3\big)\right], \end{split}\ee
and
\be \label{phi1} \ddot\phi +3\mathrm{H}\dot\phi + \left[V' - \big(12\alpha'\mathrm{H}^2+{144\beta'\mathrm{H}^4}+24\gamma' \mathrm{H}^4\big)\right] = 0. \ee
Above equations \eqref{hir1} and \eqref{phi1} are still formidably complicated to handle, and so before imposing the standard slow roll conditions, viz. $|\ddot \phi| \ll 3\mathrm{H}|\dot \phi|$ and $\dot\phi^2 \ll V(\phi)$, further simplification is required. One way is to use additional hierarchy of flow parameters \cite{SR, We1} in connection with additional degrees of freedom associated with the present model. Instead, we shall follow a completely different and unique technique. For example, redefining the potential as,

\be\label{U11} U = V -12\mathrm{H}^2(\alpha+12\mathrm{H}^2\beta+2\mathrm{H}^2\gamma),\ee
equation \eqref{phi1} takes the following form of the standard Klein-Gordon equation,

\be \label{KG1}\begin{split} \ddot\phi +3\mathrm{H}\dot\phi + U'=0 ;\end{split}. \ee
Clearly the evolution of the scalar field is driven by the re-defined potential gradient $U' = {dU\over d\phi}$, subject to the damping by the Hubble expansion $3\mathrm{H} \dot \phi$, as in the case of single field equation coupled to Einstein-Hilbert term. Note that the potential $U(\phi)$ carries all the information in connection with the coupling parameters of generalized higher order action under consideration. Further, assuming

\be\label{U12} U = V+\Lambda M^2_P - 6\mathrm{H}\dot \phi\left(\alpha' + 24 \mathrm{H}^2 \beta' + 4 \mathrm{H}^2 \gamma'\right),\ee
equation \eqref{hir1} may be reduced to the following simplified form, viz,

\be\label{Fried1} 6\alpha\mathrm{H}^2 = \frac{\dot\phi^2}{2} + U(\phi),\ee
which is essentially the non-minimally coupled Einstein's $(^0_0)$ equation. It is noteworthy that, the above two choices \eqref{U11} and \eqref{U12} of $U(\phi)$, do not contradict, since the two simply results in an evolution equation of the scalar field $\phi$, which may be different from \eqref{aphi1}, since during inflation the Hubble parameter is slowly varying. Shortly, we shall show that $\phi$ indeed falls with time, which has already been demonstrated in \cite{In1}.  Now, let us enforce the standard slow-roll conditions $\dot\phi^2\ll U$ and $|\ddot\phi|\ll 3\mathrm{H}|\dot\phi|$, so that equations (\ref{Fried1}) and (\ref{KG1}) finally reduce to,

\be\label{H2}{6\alpha}\mathrm{H}^2\simeq U, \ee
and
\be\label{Hphi} 3\mathrm{H}\dot\phi \simeq - U', \ee
respectively. Combining equations (\ref{H2}) and (\ref{Hphi}), it is possible to show that the potential slow roll parameter $\epsilon$ equals the Hubble slow roll parameter ($\epsilon_1$) under the condition,

\be\label{SR} \epsilon = - {\dot {\mathrm{H}}\over \mathrm{H}^2} = \alpha\left({U'\over U}\right)^2 - \alpha'\left({U'\over U}\right);\hspace{0.4 cm} \eta = 2 \alpha \left({U''\over U}\right).\ee
Clearly, for $\alpha =$ constant, the second term vanishes and the standard relation is restored, while the other slow-roll parameter $\eta$ remains unaltered. Further, since $\frac{\mathrm{H}}{\dot\phi}=-{U\over2\alpha U'}$ in view of equations \eqref{H2} and \eqref{Hphi}, therefore, the number of e-folds ($\mathrm{N}$) at which the present Hubble scale equals the Hubble scale during inflation, may be computed as usual in view of the following relation:

\be\label{Nphi} \mathrm{N}(\phi)\simeq \int_{t_i}^{t_f}\mathrm{H}dt=\int_{\phi_i}^{\phi_f}\frac{\mathrm{H}}{\dot\phi}d\phi = \int_{\phi_f}^{\phi_i}\Big{(}\frac{U}{2\alpha U'}\Big{)}d\phi,\ee
where, $\phi_i$ and $\phi_f$ denote the values of the scalar field at the beginning $(t_i)$ and the end $(t_f)$ of inflation. Thus, slow roll parameters reflect all the interactions, as exhibited earlier \cite{We1, In2, In4}, but here only via the redefined potential $U(\phi)$. Now, during inflation the Hubble parameter remains almost constant, and therefore while computing $U(\phi)$, one can replace it by the constant $\lambda$, without any loss of generality. Thus, using classical solutions \eqref{aphi1}, we can express \eqref{U11} as,

\be 12\mathrm{H}^2\left(\alpha+12\mathrm{H}^2\beta+2\mathrm{H}^2\gamma\right) \approx -\frac{\mathrm{H}^2\phi^2}{2},~~\mathrm{such~ that},~~ U = {1\over 2} {m^2\phi^2}-\Lambda M^2_P ,~~\mathrm{where},~~ m^2 = \lambda^2+\mathrm{H}^2 \approx 2\lambda^2.\ee
Hence, the slow roll parameters along with the number of e-folds, read as,

\begin{figure}
\begin{minipage}[h]{0.47\textwidth}
      \centering
      \begin{tabular}{|c|c|c|c|c|}
      \hline\hline
      $\alpha_0$ in ${ M^3_P}$ & $\phi_f$ in $M_P$ & $n_s$ & $r$ & $\mathrm{N}$\\
      \hline
      0.00036& 4.88810 & 0.9693 & 0.08322 & 49\\
      0.00037& 4.88822 & 0.9684 & 0.08553 & 48\\
      0.00038& 4.88833 & 0.9676 & 0.08784 & 47\\
      0.00039& 4.88845 & 0.9667 & 0.09016 & 45\\
      0.00040& 4.88856 & 0.9659 & 0.09247 & 44\\
      0.00041& 4.88868 & 0.9650 & 0.09478 & 43\\
      0.00042& 4.88879 & 0.9642 & 0.09709 & 42\\
      0.00043& 4.88890 & 0.9633 & 0.09940 & 41\\
      \hline\hline

    \end{tabular}
      \captionof{table}{Data set for the inflationary parameters taking $\phi_i=5.0 M_P$;~$m^2 = 0.084 {M^2_P}$;~$\Lambda=1 {M^2_P}$ and ~ varying $\alpha_0$, keeping $n_s$ within Planck's constraint limit.}
      \label{tab:table1}
   \end{minipage}%
   \hfill%
\begin{minipage}[h]{0.47\textwidth}
      \centering
      \begin{tabular}{|c|c|c|c|c|}
       \hline\hline
      $\alpha_0$ in $M^3_P$ & $\phi_f$ in $M_P$ & $n_s$ & $r$ & $\mathrm{N}$\\
      \hline
       0.000240& 4.88652 & 0.9795 & 0.05548 & 74\\
       0.000242& 4.88655 & 0.9793 & 0.05594 & 74\\
       0.000244& 4.88658 & 0.9792 & 0.05640 & 73\\
       0.000248& 4.88663 & 0.9788 & 0.05733 & 72\\
       0.000252& 4.88669 & 0.9785 & 0.05825 & 71\\
       0.000256& 4.88675 & 0.9782 & 0.05918 & 70\\
       0.000258& 4.88678 & 0.9780 & 0.05964 & 69\\

       \hline\hline
    \end{tabular}
      \captionof{table}{Data set for the inflationary parameters taking $\phi_i=5.0M_P$;~$m^2 = 0.084 {M^2_P}$;~$\Lambda=1{M^2_P}$ and ~ varying $\alpha_0$, keeping $r$ within Planck's constraint limit.}
      \label{tab:table2}
   \end{minipage}%
   \end {figure}
\be\label{epseta2}\epsilon = \frac{4m^4\alpha_0\phi}{(m^2\phi^2-2\Lambda M^2_P)^2}+\frac{2m^2\alpha_0}{(m^2\phi^3-2\phi\Lambda M^2_P)},\hspace{0.5 cm}\eta = \frac{4m^2\alpha_0}{m^2\phi^3-2\phi\Lambda M^2_P}.\ee
\be\label{N}\mathrm{N} = {1\over 4\alpha_0}\int_{\phi_f}^{\phi_i}{(m^2\phi^2-2\Lambda M^2_P)\over m^2}d\phi = {1\over 12\alpha_0}(\phi_i^3 - \phi_f^3)-{\Lambda M^2_P\over 2m^2\alpha_0}(\phi_i-\phi_f).\ee

Taking $\phi_i = 5 M_P$, $m^2 = 0.084 {M^2_P}$ and $\Lambda = 1 {M^2_P}$, we exhibit our data sets in a pair of tables 1 and 2. We find that inflation ends $(\epsilon = 1)$ at around $\phi_f \approx 4.49 M_P$. $\alpha_0$ is varied differently in the two tables to keep $n_s$ within the experimental limit in the first, and $r$ within the experimental limit in the second. In the first case, we see that it is not possible to reduce $r$ below $0.08$, while the second table depicts that $n_s$ exceeds the experimental limit. Of-course, the Planck's data might vary a little for different models. In this respect, the fit is fair. \\

To show the consistency of our choice of redefined potential presented in equations \eqref{U11} and \eqref{U12}, we combine the two, to obtain the following first order differential equation on $\phi$,
\be \label{Consistency} \left(\frac{\phi^3 - 6\alpha_0}{\lambda^2\phi^4-2\Lambda M^2_P\phi^2}\right) d\phi = \frac{1}{2\lambda} dt.\ee
Although the above differential equation may be solved exactly, it is extremely difficult to study its nature. We therefore neglect the second term in the numerator with respect to the first and the first term in the denominator with respect to the second, in view of our data (table-1 and table-2). Under such approximation equation \eqref{Consistency} may be expressed as:

\be  \dot\phi \approx \left[-{M_P^2\Lambda\over {\lambda\phi}}\right].\ee
Clearly, $\phi$ falls with time.\\
Additionally, it may be mentioned that the energy scale of inflation has been found to be sub-Planckian $\mathrm{H}* \approx 10^{-5} M_P$ \cite{In1}. Further, the model admits graceful exit from inflation, since the scalar field starts oscillating, $\phi \sim \pm \frac{\sqrt{\Lambda}M_P }{\sqrt{2}\lambda} \sin \left(\sqrt{2} \lambda  t-\sqrt{2}c_1 \lambda \right)$, many times over a Hubble time, driving a matter-dominated era at the end of inflation \cite{In1}.

\subsection{Case-2:}

Next, we consider an even more general action explored in \cite{In5}, which is the following,

\be\label{A2}
\begin{split}& A = \int\left[\alpha(\phi)R+\beta_1(\phi)R^2+\beta_2(\phi)\Big(R_{\mu\nu}R^{\mu\nu}-{1\over 3}R^2\Big)+\gamma(\phi)\mathcal{G}-\frac{1}{2}\phi_{,\mu}\phi^{,\mu}-V(\phi)\right] d^{4}x \sqrt{-g}.\end{split} \ee
Note that here we consider the additional curvature squared term, viz. $R_{\mu\nu}^2$, with an additional $\phi$ dependent coupling parameter $\beta_2(\phi)$.

\subsubsection{Field equations and classical solutions:}

Due to diffeomorphic invariance (energy constraint), only two components of Einstein's equations are independent, as mentioned previously. We therefore consider the $(^0_0)$ and the $\phi$ variation equations in the background of Robertson-Walker metric \eqref{RW}, which are,

\be \label{002} \begin{split} -\frac{6\alpha}{a^2}\bigg{(}\dot a^2+ k\bigg{)}-\frac{6\alpha' \dot a\dot\phi}{a} &-36\beta_1 \bigg{(}\frac{2\dot a\dddot a}{a^2}-\frac{\ddot a^2}{a^2} +\frac{2\dot a^2 \ddot a}{a^3}-\frac{3\dot a^4}{a^4}-{2k\dot a^2\over a^4}+{k^2\over a^4}\bigg{)}-72\beta_1'\dot\phi\bigg{(}\frac{\dot a\ddot a}{a^2}+\frac{\dot a^3}{a^3}+{k\dot a\over a^3}\bigg{)}\\&+{6\beta_2'\dot\phi}\bigg({2\dot a^3\over a^3}+{3k\dot a\over a^3} \bigg)-24\gamma'\dot\phi\bigg{(}{\dot a^3\over a^3}+{k\dot a\over a^3}\bigg{)}+\bigg{(}\frac{\dot\phi^2}{2}+V \bigg{)}=0\end{split} \ee
and
 \be \label{phivariation2} \begin{split} -6\alpha'\bigg{(}a^2\ddot a+a\dot a^2+ka\bigg{)}& -36\beta_1'\bigg{(}a\ddot a^2+2\dot a^2\ddot a+ \frac{\dot a^4}{a}+{k^2\over a}+{2k\dot a^2\over a}+2k\ddot a\bigg{)} +12\beta_2'\bigg({\dot a^2\ddot a}+{k\ddot a} \bigg)\\&+3a^2\dot a \dot\phi -24\gamma'\bigg{(}\dot a^2\ddot a+k\ddot a\bigg{)} +a^3\bigg{(}\ddot\phi+V'\bigg{)}=0 .\end{split}\ee
The above field equations also admit the following de-Sitter solution in the spatially flat space ($k=0$),

\be\label{aphi2} \begin{split} a = &a_0e^{\lambda t}; ~~~ \phi= \phi_0e^{-\lambda t},~~\mathrm{under~ the~ condition},\\&
\alpha(\phi) = {\alpha_0\over \phi};~~ V(\phi) = \frac{1}{2} \lambda^2 \phi^2;~~\mathrm{and}~~ \beta_2-2\Big(6 \beta_1 +  \gamma\Big) = {1\over \lambda^2}\left({\alpha_0\over \phi} +{\phi^2\over 24}\right),\end{split}\ee
where, $a_0$, $\phi_0$, $\alpha_0$ and $\lambda$ are arbitrary constants while $\beta_1(\phi)$, $\beta_2(\phi)$ and $\gamma(\phi)$ remain arbitrary functions of $\phi$ and are related through equation \eqref{aphi2}, after setting the constant of integration to zero without any loss of generality. However, for canonical quantization, arbitrariness should be removed, since one is required to order the operators. In \cite{In5} we therefore removed the arbitrariness on $\beta$ and $\gamma$, following a simple assumption viz,

\be\label{const} \beta_1 = \frac{\alpha_0}{12\lambda^2\phi} = {\beta_{01}\over \phi};\hspace{0.2 cm}\beta_2 = \frac{2\alpha_0}{\lambda^2\phi} = {\beta_{02}\over \phi};\hspace{0.2 cm}\gamma = -\frac{\phi^2}{48\lambda^2} = {\gamma_0\phi^2}\hspace{0.2 cm} \mathrm{where},\hspace{0.2 cm}\beta_{01} = {\alpha_0\over 12 \lambda^2};\hspace{0.2 cm}\beta_{02} = {2\alpha_0\over \lambda^2};~~\gamma_0 = -{1\over 48\lambda^2},\ee
where, $\beta_{01}$, and $\beta_{02}$ are constants.

\subsubsection{Inflation under Slow Roll Approximation:}

As before, we express \eqref{002} and \eqref{phivariation2} in terms of Hubble parameter for spatially flat space $(k = 0)$ as,

\be \label{hir2}\begin{split} 6\alpha \mathrm{H}^2 = \frac{\dot\phi^2}{2} + \left[V - \big(6\alpha'\dot\phi\mathrm{H} + 144\beta'_1\dot\phi\mathrm{H} ^3 -12\beta'_2{\dot\phi} \mathrm{H}^3 + 24\gamma'\dot\phi \mathrm{H}^3\big)\right], \end{split}\ee
and
\be \label{phi2} \ddot\phi +3\mathrm{H}\dot\phi + \left[V' - \big(12\alpha'\mathrm{H}^2+{144\beta'_1\mathrm{H}^4}-12\beta'_2 \mathrm{H}^4 +24\gamma' \mathrm{H}^4\big)\right] = 0, \ee
respectively. Here again, instead of using additional hierarchy of flow parameters \cite{SR, We1}, we define a potential in the following manner:

\be\label{U21} U = V -12\mathrm{H}^2(\alpha+12\mathrm{H}^2\beta_1-\mathrm{H}^2\beta_2+2\mathrm{H}^2\gamma).\ee
so that equation (\ref{phi2}) takes the following standard form of Klein-Gordon Equation,

\be \label{KG2}\begin{split} \ddot\phi +3\mathrm{H}\dot\phi + U'=0. \end{split} \ee
Clearly as before, the evolution of the scalar field is driven by the re-defined potential gradient $U' = {dU\over d\phi}$, subject to damping by the Hubble expansion $3\mathrm{H} \dot \phi$, as in the case of single field equation. Further, the potential $U(\phi)$ carries all the information in connection with the coupling parameters of generalised higher order action under consideration. Further assuming,

\be\label{U22} U = V - 6\mathrm{H}\dot \phi\left(\alpha' + 24 \mathrm{H}^2 \beta'_1-2\mathrm{H}^2 \beta'_2+ 4 \mathrm{H}^2 \gamma'\right),\ee
equation \eqref{hir2} may be reduced to the following simplified form, viz,

\be\label{Fried2} 6\alpha\mathrm{H}^2 = \frac{\dot\phi^2}{2} + U(\phi),\ee
which is simply the Friedmann equation with a single scalar field and non-minimal coupling $\alpha(\phi)$. Here again, the two choices on the redefined potential $U(\phi)$ made in \eqref{U21} and \eqref{U22}, do not confront in any case, since the combination simply gives the evolution equation of the scalar field.
During slow roll, the Hubble parameter $\mathrm{H}$ almost remains unaltered. Thus replacing $\mathrm{H}$ by $\lambda$, and using the forms of the parameters $\alpha(\phi)$ presented in \eqref{aphi2}, along with $\beta_1(\phi)$, $\beta_2(\phi)$ and $\gamma(\phi)$ assumed in \eqref{const}, the two relations \eqref{U21} and \eqref{U22} lead to the following first order differential equation on $\phi$,

\be \label{Consistency2} \left(\frac{\phi^3 - 6\alpha_0}{\phi^4}\right) d\phi = \frac{\lambda}{2} dt,\ee
which can immediately be integrated to yield,

\be \label{phidecay2}\ln{\phi} + \frac{2\alpha_0}{\phi^3} = \frac{\lambda}{2}(t - t_0).\ee
Clearly, if $\phi$ is not too large, $\ln{\phi}$ remains subdominant, and $\phi$ falls-of with time, as expected during inflationary regime.\\

 Now, under slow roll approximation $(\dot \phi^2 \ll U(\phi)~ \mathrm{and}~ \ddot \phi << 3\mathrm{H}|\dot\phi|)$, the effective Friedmann \eqref{Fried2} and the Klein-Gordon \eqref{KG2} equations take the same form of equations \eqref{H2} and \eqref{Hphi}. Hence $\frac{\mathrm{H}}{\dot\phi}=-{U\over2\alpha U'}$, as before. Therefore, the slow roll parameters and the number of e-folds, at which the present Hubble scale equals the Hubble scale during inflation, may be computed as:

\be\label{SR2} \epsilon = - {\dot {\mathrm{H}}\over \mathrm{H}^2} = \alpha\left({U'\over U}\right)^2 - \alpha'\left({U'\over U}\right);\hspace{0.4 cm} \eta = 2 \alpha \left({U''\over U}\right).\ee

\be\label{Nphi2} \mathrm{N}(\phi)\simeq \int_{t_i}^{t_f}\mathrm{H}dt=\int_{\phi_i}^{\phi_f}\frac{\mathrm{H}}{\dot\phi}d\phi\simeq \int_{\phi_f}^{\phi_i}\Big{(}\frac{U}{2\alpha U'}\Big{)}d\phi,\ee
where, $\phi_i$ and $\phi_f$ denote the values of the scalar field at the beginning $(t_i)$ and the end $(t_f)$ of inflation. Thus, slow roll parameters reflect all the interactions, as exhibited earlier \cite{We1, In2, In4}, via the redefined potential $U(\phi)$. Now, let us make the following choice of the redefined potential,

\be U = {1\over 2} {m^2\phi^2}-u_0,~~\mathrm{where},~~ m^2 = \lambda^2+\mathrm{H}^2 \approx 2\lambda^2,\ee
where $u_0$ is a constant which is essentially the vacuum energy density, i.e. the cosmological constant, that we omitted from the action. Thus, the slow roll parameters $\epsilon$ and $\eta$ \eqref{SR2} and the number of e-folding $\mathrm{N}$ \eqref{Nphi2} take the following forms,

\be\label{epseta2}\epsilon = \frac{4m^4\alpha_0\phi}{(m^2\phi^2-2u_0)^2}+\frac{2m^2\alpha_0}{(m^2\phi^3-2 u_0 \phi )},\hspace{0.5 cm}\eta = \frac{4m^2\alpha_0}{m^2\phi^3-2 u_0 \phi  }.\ee
\be\label{N}\mathrm{N} = {1\over 4\alpha_0}\int_{\phi_f}^{\phi_i}{(m^2\phi^2-2 u_0 )\over m^2}d\phi = {1\over 12\alpha_0}(\phi_i^3 - \phi_f^3)-{u_0\over 2m^2\alpha_0}(\phi_i-\phi_f).\ee
Here again we present two sets of data in table 3 and table 4 taking $\phi_i=1.54 M_P$;~$m^2 = 0.9 {M^2_P}$;~$u_0=1 {M^4_P}$, so that inflation ends $(\epsilon = 1)$ around $\phi_f \approx 1.49 M_P$. In table 3, we have varied $\alpha_0$ in such a manner $(1.80\times 10^{-5} M_P^3<\alpha_0 < 2.15\times 10^{-5}M_P^3)$ that the scalar tilt , i.e. the spectral index lie very much within the specified range, i.e. $0.964 < n_s < 0.970$. The number of e-folds $ 42\le N \le 50$ is enough to solve the horizon and the flatness problem. However, the tensor to scalar ratio does not admit value $r < 0.06$. On the contrary, in table 4, we have kept the tensor to scalar ratio within the specified limit $r < 0.06$, and find that the spectral index goes beyond experimental limit. Further, the number of e-folds becomes a bit large. Since, Planck's data has been analysed following a particular model, so some deviation is expected for more involved models, under present consideration. In this case also it is important to mentioned that, the energy scale of inflation has been found to be sub-Planckian $(\mathrm{H}*) \approx 10^{-5} M_P$ \cite{In5}. The model also admits graceful exit from inflation, since the scalar field starts oscillating $(\phi \sim e^{i\sqrt 2\lambda t})$ many times over a Hubble time, driving a matter-dominated era at the end of inflation \cite{In5}.

\begin{figure}
\begin{minipage}[h]{0.47\textwidth}
      \centering
      \begin{tabular}{|c|c|c|c|c|}
      \hline\hline
      $\alpha_0$ in $\times10^{-5}M^3_P$ & $\phi_f$ in $M_P$ & $n_s$ & $r$ & $\mathrm{N}$\\
      \hline
       1.80& 1.49419 & 0.9699 & 0.08201 & 50\\
       1.85& 1.49424 & 0.9690 & 0.08429 & 49\\
       1.90& 1.49429 & 0.9682 & 0.08657 & 48\\
       1.95& 1.49433 & 0.9674 & 0.08884 & 47\\
       2.00& 1.49438 & 0.9665 & 0.09112 & 46\\
       2.05& 1.49442 & 0.9657 & 0.09340 & 44\\
       2.10& 1.49447 & 0.9649 & 0.09568 & 43\\
       2.15& 1.49451 & 0.9640 & 0.09796 & 42\\
    \hline\hline
    \end{tabular}
      \captionof{table}{Data set for the inflationary parameters taking $\phi_i=1.54 M_P$;~$m^2 = 0.9 {M^2_P}$;~$u_0=1 {M^4_P}$ and ~ varying $\alpha_0$, keeping $n_s$ within Plank constraint limit.}
      \label{tab:table3}
   \end{minipage}%
   \hfill%
\begin{minipage}[h]{0.47\textwidth}
      \centering
      \begin{tabular}{|c|c|c|c|c|}
      \hline\hline
      $\alpha_0$ in $\times10^{-5}{ M^3_P}$ & $\phi_f$ in $M_P$ & $n_s$ & $r$ & $\mathrm{N}$\\
      \hline
      1.20& 1.49355 & 0.9799 & 0.05467 & 76\\
      1.22& 1.49358 & 0.9796 & 0.05558 & 75\\
      1.24& 1.49360 & 0.9792 & 0.05650 & 73\\
      1.26& 1.49362 & 0.9789 & 0.05740 & 72\\
      1.28& 1.49365 & 0.9786 & 0.05832 & 71\\
      1.30& 1.49367 & 0.9782 & 0.05923 & 70\\
    \hline\hline
    \end{tabular}
      \captionof{table}{Data set for the inflationary parameters taking $\phi_i=1.54M_P$;~$m^2 = 0.9 {M^2_P}$;~$u_0=1{M^4_P}$ and ~ varying $\alpha_0$, keeping $r$ within Plank constraint limit.}
      \label{tab:table4}
   \end{minipage}%
   \end {figure}

\subsection{Case-3}

Although, Gauss-Bonnet term is constructed from higher order curvature invariant terms, the beauty lies in the fact that, it does not contain anything above second derivative, and hence is free from ghost degrees of freedom and also renormalizable. The problem is, it suffers from the pathology of `Branched Hamiltonian'\cite{BH,I6}. The presence of cubic kinetic term and quadratic constraints appearing through Gauss-Bonnet combination, makes the theory intrinsically nonlinear. Even its linearized version is cubic rather than quadratic. Since, the expression for velocities are multi-valued functions of momentum, it results in the so called multiply branched Hamiltonian with cusps. This makes classical solution unpredictable, as at any instant of time, one can jump from one branch of the Hamiltonian to the other. Further, the momentum does not provide a complete set of commuting observable, resulting in non-unitary time evolution of quantum states. Such a pronounced exotic behaviour does not allow Hamiltonian formulation following conventional Legendre transformation. There is no unique resolution to this issue. However, it was shown that the pathology may be bypassed by adding curvature squared term \cite{BH,I6}. Let us therefore consider the action as considered earlier in \cite{I6} which is,

\be \label{A3} A = \int \left[{R \over 16\pi G} +\xi(\phi)\left(\mathcal{G} + \beta R^2 \right)- {1\over 2}\phi_{,\mu}\phi^{,\mu} - V(\phi)\right]\sqrt{-g}~d^4x.\ee
Note that we have omitted scalar coupling with Einstein-Hilbert sector and introduced the same coupling parameter $\xi(\phi)$ with the Gauss-Bonnet and the $R^2$ term.

\subsubsection{Field equations and classical solutions:}

The $(^0_0)$ and the $\phi$ variation equations are,

\be\label{300}\begin{split}{{\dot a}^2\over a^2}&=-96\beta\xi\pi G\left[\frac{2\dot a\dddot a}{a^2}-\frac{\ddot a^2}{a^2}+\frac{2\dot a^2\ddot a}{a^3}-\frac{3\dot a^4}{a^4}\right]-192\beta\pi G\xi'{\dot\phi}\left(\frac{\dot a\ddot a}{a^2}+\frac{\dot a^3}{a^3}\right)\\&\hspace{1in}-64\pi G\xi'{\dot\phi}\left(\frac{\dot a^3}{a^3}\right)+{8\pi G\over 3}\left({{\dot\phi^2}\over 2}+V \right).\end{split}\ee
\be\label{3phivar} -24\xi'\dot a^2\ddot a-36\beta\xi'a\ddot a^2-72\xi'\dot a^2\ddot a-36\beta\xi'\frac{\dot a^4}{a}+3a^2\dot a\dot\phi+a^3\left(\ddot\phi+V'\right)=0.\ee
If we now seek classical de-Sitter solution in the form

\be\label{aphi3}a = a_0e^{\lambda t}$ and $\phi=\phi_0 e^{-\lambda t},\ee
 then the coupling parameter and the potential are fixed as
\be\label{coup3}\xi(\phi)=\xi\phi^{-2},\mathrm{and}~~V(\phi)=V_1+V_0\phi^2,\ee
restricting the constants to $V_1= \frac{3\lambda^2}{8\pi G},~~\beta=-{1\over 6}$, and $V_0=-\frac{\lambda^2}{2}$, where $\xi$ and $\lambda$ are constant.

\subsubsection{Inflation under Slow Roll approximation:}

Note that during inflation the Hubble parameter varies slowly and hence we can replace the constant $\lambda$ by $\mathrm{H}$, without loss of generality. Here, instead of redefining the potential as in the previous two cases, we consider an additional slow roll parameter viz. $\delta_1 = 4\mathrm{H}{\dot\xi } \ll1$, following the hierarchy of flow parameters \cite{SR, We1}. Thus we have three slow roll parameters $\epsilon,~\eta,~\delta_1$ at hand, in view of which \eqref{300} and \eqref{3phivar} may be approximated to \cite{I6},

\be\label{SR} \mathrm{H}^2 \simeq {1\over 3 M^2_P}V,\hspace{0.5in}\text{and}\hspace{0.5in} \mathrm{H}{\dot\phi}\simeq -{1\over 3}V\mathcal{Q},\ee
where $\mathcal{Q}= {V'\over V}$. Now, in view of the above form of a monomial potential and an inverse monomial GB coupling \eqref{coup3}, namely
$V(\phi)=V_1+V_0\phi^2$ and~~$\xi(\phi)=\xi\phi^{-2}$, where $V_0, V_1, \xi$ are constants, the slow roll parameters and the number of e-folds may be expressed as,

\be\label{epseta3}\begin{split}&\epsilon = \frac{2\phi^2M_P^2}{({V_1\over V_0}+\phi^2)^2},\hspace{0.5 cm}\eta = \frac{2V_0}{({V_1\over V_0}+\phi^2)},\\&
\mathrm{N} = {1\over 2 M_P^2}\int_{\phi_f}^{\phi_i}{(V_1+V_0\phi^2)\over V_0\phi} d\phi = {1\over 2M_P^2}\left[{V_1\over V_0}ln({\phi_i\over \phi_f})+{(\phi_i^2-\phi_f^2)\over2}\right].\end{split}\ee

\begin{center}
\begin{minipage}[h]{0.47\textwidth}
      \centering
\begin{tabular}{|c|c|c|c|c|}
\hline\hline
 ${V_1\over V_0}~in M_P^2$ &  $\phi_f ~in M_P$  & $r$ & $n_s$ & $\mathrm{N}$ \\\hline
  -6.0 & 3.2566 & 0.1244 &.9685 & 60 \\\hline
  -5.5 &3.1566 & 0.1240  &.9687 & 60\\\hline
  -5.0 &3.0523 & 0.1235 &.9688 & 61 \\\hline
 -4.5 & 2.9432 & 0.1231 &.9690 & 61 \\\hline
 -4.0 & 2.8284 & 0.1226 &.9691 & 62 \\\hline
 -3.5 & 2.7071 & 0.1221 &.9693 & 62 \\\hline
 -3.0 & 2.5780 & 0.1217 &.9694 & 63 \\\hline
 -2.5 & 2.4392 & 0.1212 &.9696 & 64 \\\hline
 -2.0 & 2.2883 & 0.1208 &.9697 & 64 \\\hline
 -1.5 & 2.1213 & 0.1203 &.9698 & 65 \\\hline
 -1.0 & 1.9319 & 0.1199 &.9699 & 65 \\\hline
\hline
\end{tabular}
      \captionof{table}{${\phi_i}=16.4M_P,V_0=1 M_P^2$}
      \label{table:5}
   \end{minipage}%
   \end{center}
In table 5 we present a data set under the choice ${\phi_i}=16.4M_P ~\mathrm{and} ~V_0=1 M_P^2$, while ${V_1\over V_0}$ is varied within the range $-6.0 M_P^2 < {V_1\over V_0} < -1.0M_P^2$. Although the spectral index of scalar perturbation lie within the experimental limit $(0.96 \le n_s \le 0.97)$ and the number of e-folding ranges within $60 < N < 65$, which is sufficient to solve the horizon and flatness problems, it has not been possible to keep the tensor to scalar ratio $(r > 0.1)$ within the observational limit. Further, even though the model allows graceful exit from inflation, since the scalar field exhibit oscillatory behaviour  as $\Big(\phi \sim \sqrt{V_1\over V_0} \sin \left(\sqrt{2} t \sqrt{V_0}\pm c_1 \sqrt{V_0}\right)\Big)$ at the end of inflation; however the energy scale of inflation is super Planckian $\mathrm{H}* \approx 9.38 M_P$. Nonetheless, before discarding this model, we need to apply redefined potential technique, instead of the additional slow roll parameter $\delta_1$, which we pose in the future.

\subsection{Case-4:}

In this subsection, we shall consider yet another higher-order modified gravitational action, which has not been treated earlier. Gauss-Bonnet term being topologically invariant, does not contribute to the field equation, as already mentioned. Therefore a dilatonic (scalar) coupling is necessary. Recently, $F(\mathcal{G})$ theory has been proposed as an alternative to the dark energy. It is interesting to note that different powers (other than one) of the Gauss-Bonnet term $\mathcal{G}$, may be incorporated in the action without dilatonic coupling. It has been found that a typical form of $F(\mathcal{G}) = F_0 \mathcal{G}^m + F_1 \mathcal{G}^n$, might unify early inflation with the late-time cosmic acceleration \cite{Cog}. Particularly for $m < {1\over 2}$, late-time acceleration may be addressed, while for $n > 1$, early inflation is admissible. Since we are interested in the evolution of the early universe, so we leave the first term and choose $n = 2$, for simplicity, and express the action in the presence of a Gauss-Bonnet dilatonic term, which as mentioned is an outcome of weak field approximation of different versions of string theory, as:

\be \label{A4} A = \int \left[\alpha(\phi)(R - 2\Lambda) +\beta(\phi)\mathcal{G} + \gamma\mathcal{G}^2 - {1\over 2}\phi_{,\mu}\phi^{,\mu} - V(\phi)\right]\sqrt{-g}~d^4x.\ee

\subsubsection{Field equations and classical solutions:}

The $(^0_0)$ and the $\phi$ variation equations in connection with the above action \eqref{A4} in the background of Robertson-Walker metric \eqref{RW} are the following:

\be\label{004}\begin{split}\\& 2\alpha \left({3\dot a^2\over a^2}-\Lambda\right) + 18\gamma\left[64\left({3\dot a^6 \ddot a\over a^7} + {\dot a^5 \dddot a\over a^6}\right) +96 \left({\dot a^4\over  a^4}+{\dot a^2\ddot a\over  a^3}\right)-576\left({\dot a^8\over  a^8}+{\dot a^6\ddot a\over  a^7}\right) + {480 \dot a^8\over  a^8}\right]\\&+{6\alpha'\dot\phi \dot a\over a}+{24\beta'\dot\phi \dot a^3\over a^3} = {1\over 2}\dot\phi^2 + V(\phi).\end{split}\ee
\be\label{phivariation4} \begin{split} \\& \ddot\phi + {3}{\dot a\over a}\dot\phi + V'-6\alpha'\left({\dot a^2\over a^2}+{\ddot a\over a}\right)+2\Lambda\alpha'-24\beta'\left({\dot a^4 \over a^4}+{\dot a^2\ddot a\over a^3}\right)+{{24\beta'\dot a^4}\over a^4} = 0.\end{split}\ee
Now seeking inflationary solution of the above classical field equations in the following standard de-Sitter form,
\be\label{aphi4} a = a_0e^{\lambda t}; ~~~ \phi= \phi_0e^{-\lambda t},\ee
the parameters $\alpha$, $\beta$ and potential $V(\phi)$ are fixed as,

\be\label{param4}\begin{split}& \alpha(\phi) = {\alpha_0\over \phi};\hspace{0.2 in}\beta(\phi) = -{\phi^2\over 48\lambda^2}-{\alpha_0\over \phi}\left({1\over 2\lambda^2}-{\Lambda \over 12\lambda^4}\right)=-{{\alpha_0 \beta_0}\over \phi}-{\beta_1 \phi^2};\\&
 V(\phi) = \frac{1}{2} \lambda^2 \phi^2-576\gamma\lambda^8;\hspace{0.2 in} \beta _0=\left({1\over 2\lambda^2}-{\Lambda \over 12\lambda^4}\right);\hspace{0.2 in}\beta_1={1\over 48\lambda^2}; \hspace{0.2 in}\gamma=\gamma_0,\end{split}\ee
where, $a_0$, $\alpha_0$, $\gamma_0$, $\phi_0$, and $\lambda$ are arbitrary constants, while $\beta_0$, $\beta_1$ are related though $\lambda$.\\

\subsubsection{Inflation under Slow Roll Approximation:}

As before, let us express equations (\ref{004}) and (\ref{phivariation4}) in terms of the Hubble parameter $\mathrm{H}$ in the spatially flat space $(k = 0)$ respectively as,

\be\label{hir4} 6\alpha \mathrm{H}^2=576\gamma_0{\mathrm{H}^8}-6\alpha'\dot\phi\mathrm{H}-24\beta'\dot\phi\mathrm{H}^3+\bigg{(}{V}+2\Lambda\alpha\bigg{)}+{{\dot\phi}^2\over 2},\ee
and
\be\label{phi4}\ddot\phi+ 3\mathrm{H}\dot\phi =-V'-2\alpha'\Lambda+12\alpha'\mathrm{H}^2+24\beta'{\mathrm{H}^4}. \ee
As in case-1 and case-2, here again we reduce the above set of highly complicated equations by redefining the potential, instead of using the additional hierarchy of flow parameters \cite{SR, We1}. For example, choosing the potential as,

\be\label{4U1} U = V +2\alpha\Lambda-12\mathrm{H}^2(\alpha+2\mathrm{H}^2\beta).\ee
The above consideration again modifies the equation (\ref{phi4}) to the standard form of Klein-Gordon Equation as,

\be\label{KG4} \ddot\phi +3\mathrm{H}\dot\phi + U'=0.\ee
Further, assuming

\be\label{4U2} U = V + 2\alpha\Lambda +576\gamma_0\mathrm{H}^8 - 6\mathrm{H}\dot \phi\left(\alpha' + 4 \mathrm{H}^2 \beta'\right),\ee
equation \eqref{hir4} may also be reduced to the following simplified form, viz,
\be\label{Fried4} 6\alpha\mathrm{H}^2 = \frac{\dot\phi^2}{2} + U(\phi).\ee
Here again we mention that, the two choices of the redefined potential $U(\phi)$ made in \eqref{4U1} and \eqref{4U2}, do not contradict each other, rather equating the two redefined potential, one obtains the following first order differential equation on $\phi$ (since the Hubble parameter being slowly varying, may be treated almost as a constant):
\be \label{Consistency4} \dot\phi (t) = {2\mathrm{H}\alpha + 4\mathrm{H}^3 \beta+96\gamma_0\mathrm{H}^7 \over \alpha' + 4 \mathrm{H}^2\beta'}.\ee
Shortly, the behaviour of $\phi$ with time will be exhibited. We now enforce the standard slow-roll conditions $\dot\phi^2\ll U$ and $|\ddot\phi|\ll 3\mathrm{H}|\dot\phi|$, on equations (\ref{Fried4}) and (\ref{KG4}), which thus finally reduce to,

\be\label{4H2}{6\alpha}\mathrm{H}^2\simeq U, \ee
and
\be\label{4Hphi} 3\mathrm{H}\dot\phi \simeq - U',\ee
respectively. Let us now compute the functional form of $U = U(\phi)$. For this purpose, we consider the same quadratic  form of the potential as, $V(\phi) = {1\over 2}\lambda^2\phi^2-576\gamma_0\lambda^8$, along with given forms of $\alpha(\phi)$, $\beta(\phi)$ in \eqref{param4}, which satisfy classical de-Sitter solutions. As already mentioned, during inflation the Hubble parameter remains almost constant, and therefore while computing $U(\phi)$, one can replace it by the constant $\mathrm{H} \approx \lambda$, without any loss of generality. Thus from \eqref{4U1} one obtains,

\be\label{Uphi4} U = {\lambda^2\phi^2}-576\gamma_0\lambda^8={m^2\phi^2}- C_0,\ee
where $m$ may be treated as the mass of the scalar field and $C_0 = 576\gamma_0\lambda^8$. Now, for the above form of $U(\phi)$ \eqref{Uphi4}, the slow roll parameters read as,

\be\label{epseta4}\begin{split}&\epsilon = \frac{4m^4\alpha_0\phi}{(m^2\phi^2-C_0)^2}+\frac{2m^2\alpha_0}{(m^2\phi^3-\phi C_0)},\hspace{0.5 cm}\eta = \frac{4m^2\alpha_0}{m^2\phi^3-\phi C_0},\\&
\mathrm{N} = {1\over 4\alpha_0}\int_{\phi_f}^{\phi_i}{(m^2\phi^2-C_0)\over m^2}d\phi = {1\over 12\alpha_0}(\phi_i^3 - \phi_f^3)-{C_0\over 8m^2\alpha_0}(\phi_i-\phi_f).\end{split}\ee
\begin{figure}

   \begin{minipage}[h]{0.47\textwidth}
      \centering
      \begin{tabular}{|c|c|c|c|c|}
      \hline\hline
      $\alpha_0$ in $M^3_P$ & $\phi_f$ in $M_P$ & $n_s$ & $r$ & $\mathrm{N}$\\
      \hline
      0.0076& 2.61173 & 0.9668 & 0.0992 & 51\\
      0.0075& 2.61207 & 0.9673 & 0.0979 & 52\\
      0.0074& 2.61242 & 0.9677 & 0.0966 & 53\\
      0.0073& 2.61277 & 0.9681 & 0.0953 & 54\\
      0.0072& 2.61312 & 0.9686 & 0.0940 & 54\\
      0.0071& 2.61348 & 0.9690 & 0.0927 & 55\\
      0.0070& 2.61383 & 0.9694 & 0.0914 & 56\\

       \hline\hline
    \end{tabular}
      \captionof{table}{Data set for the inflationary parameters taking $\phi_i=3.40 M_P$;~$m^2 = 1\times10^{-10} {M^2_P}$;~$C_0=7.10\times10^{-10} {M^4_P}$ and varying $\alpha_0$, keeping $n_s$ within Planck's constraint limit.}
      \label{tab:table6}
   \end{minipage}%
   \hfill%
\begin{minipage}[h]{0.47\textwidth}
      \centering
      \begin{tabular}{|c|c|c|c|c|}
      \hline\hline
      $\alpha_0$ in ${ M^3_P}$ & $\phi_f$ in $M_P$ & $n_s$ & $r$ & $\mathrm{N}$\\
      \hline
      0.0081& 2.46388 & 0.9738 & 0.0796 & 67\\
      0.0080& 2.46423 & 0.9741 & 0.0786 & 68\\
      0.0079& 2.46457 & 0.9745 & 0.0776 & 69\\
      0.0078& 2.46492 & 0.9748 & 0.0766 & 69\\
      0.0077& 2.46527 & 0.9751 & 0.0756 & 70\\
      0.0076& 2.46562 & 0.9754 & 0.0747 & 71\\
      0.0075& 2.46598 & 0.9754 & 0.0737 & 72\\
        \hline\hline
    \end{tabular}
      \captionof{table}{Data set for the inflationary parameters taking $\phi_i=3.40 M_P$;~$m^2 = 1\times10^{-10} {M^2_P}$;~$C_0=6.35\times10^{-10}{M^4_P}$ and varying $\alpha_0$ in such a manner that $r$ is small.}
      \label{tab:table7}
   \end{minipage}%
\end{figure}

Here again, we compute the inflationary parameters taking $\phi_i=3.40 M_P$ and $m^2 = 1\times10^{-10} {M^2_P}$. In table 6 we consider $C_0=7.10\times10^{-10} {M^4_P}$ and vary $\alpha_0$ within the range $ 0.0070 M_P^3< \alpha_0 < 0.0076M_P^3$, so that the spectral index lies within the experimental limit $0.966 < n_s <0.970$. Although the number of e-folds remain within the range $51 < N < 56$, which is sufficient to solve the horizon and flatness problems, the tensor to scalar ratio can not be reduced below $r = 0.09$. Therefore, although $r$ fits fairly well with the Planck's data \cite{Planck1, Planck2}, it does not fall within the range specified by other experiments, viz, BAO, BICEP, BK15 keck Array data. In table 7, we fix $C_0=6.35\times10^{-10}{M^4_P}$ and vary $\alpha_0$ within the range $0.0075 M_P^3 < \alpha_0 < 0.0081M_P^3$. As a result, the specified range of spectral index is relaxed. Nonetheless, it is still not possible to keep $r < 0.07$.\\

Let us therefore proceed to find the energy scale of inflation. In view of the above form of $U(\phi)$ \eqref{Uphi4}, we obtain the following expression from equation\eqref{Fried4},
\be \label{HM}6{\alpha_0\over\phi}\mathrm{H}^2= m^2\phi^2-C_0.\ee
Now, if we choose the value of $C_0 = 7.10\times 10^{-10} M^4_P$, together with a value of $\alpha_0 =0.0075 M^3_P$,  with, $m^2 = 1\times10^{-10} {M^2_P},~\phi_i=3.40 M_P$, as depicted in the table-6, we simply find,

\be \mathrm{H}^2=\frac{(m^2\phi^3-C_0\phi)}{6\alpha_0}, \hspace{0.3in} \mathrm{and~hence}, \hspace{0.3in}  \mathrm{H} ^2 \approx 3.37\times10^{-8} M^2_P,\ee
 Therefore, the energy scale of inflation has been found to be sub-Planckian $(\mathrm{H}* \approx 10^{-4} M_P)$.\\

 To exhibit consistency of our choice of redefined potential presented in \eqref{param4}, and treating the Hubble parameter to be nearly constant during inflation $\mathrm{H} \approx \lambda$, the equation \eqref{Consistency4} may now be expressed as,

\be\label{soln4}\dot\phi (t)={\lambda \phi\big[2C_0\phi+4\Lambda \alpha_0 - \lambda^2 \phi^3\big]\over 4\alpha_0\big(3\lambda^2 - \Lambda\big) - 2\lambda^2\phi^3}.\ee
The above differential equation cannot be integrated analytically. Nonetheless, since  $C_0\approx 10^{-10}M_p^4$ and $\lambda^2\approx 10^{-8}M_P^2$, the terms associated with these parameters may be neglected from the numerator and denominator. As a result \eqref{soln4} may be suitably approximated to,

\be{\dot\phi (t)}=-\left[{{\lambda \phi(4\alpha_0\Lambda)}\over {4\alpha_0\Lambda}}\right],\hspace{1cm} \mathrm{that~is} \hspace{1cm} \dot \phi = -{\lambda \phi}. \ee
This is a remakable outcome, since even after making an additional assumption of redefined potential, the scalar field evolves identically as in the classical de-Sitter solution \eqref{aphi4}. Having shown that $\phi$ decays, let us now express equation \eqref{Fried4} as,

\be \label{Hubble} {3\mathrm{H}^2\over m^2} = {\phi\over 2\alpha_0}\left({\dot\phi^2\over 2m^2} + \phi^2-{C_0\over m^2}\right).\ee
Note that for single scalar field, the above equation reads as: $3\mathrm{H}^2 = {1\over 2M_p^2}(\dot\phi^2 + 2 m^2 \phi^2-2C_0)$. Since at the end of inflation, ${\phi\over 2\alpha_0} \sim {226 M_p^{-2}}$, according to the present data set, so once the Hubble rate ($\mathrm{H}$) falls below $m$, this equation \eqref{Hubble} may be approximated to,

\be \dot\phi^2 \approx -2(m^2 \phi^2- C_0),\ee
which may immediately be integrated to yield,

\be\phi (t)=\pm \frac{\sqrt{C_0} \tan \left[m (\sqrt{2} t-t_0)  \right]}{m  \sqrt{\tan ^2\left[ m  (\sqrt{2}t-t_0) \right]+1}},\ee
and may further be simplified to obtain
\be\phi (t)=\pm \frac{\sqrt{C_0}}{m}\sin \left[m (\sqrt{2} t-t_0)\right]. \ee
Where $t_0$ is the constant of integration. Thus the scalar field starts oscillating many times over a Hubble time, driving a matter-dominated era as  inflation ends.

\subsection{Case-5:}

All the modified theories of gravity considered so far are higher order theories. There is yet another class of modified theories, viz, the non-minimally coupled scalar tensor theories of gravity. Such theories are essentially dark energy quintessence models. Here we consider a pure ( having regular kinetic energy term and being devoid of higher-order terms) non-minimally coupled scalar-tensor theory of gravity, as considered in\cite{Beh,Sym} for which the action is expressed in the form,

\be \label{2.1} A = \int \left[f(\phi) R - {\omega(\phi)\over \phi}\phi_{,\mu}\phi^{^,\mu} - V(\phi) - \mathcal{L}_m \right]\sqrt{-g} d^4 x, \ee
where, $\mathcal{L}_m$ is the matter Lagrangian density, $f(\phi)$ is the coupling parameter, while, $\omega(\phi)$ is the variable Brans-Dicke parameter.

\subsubsection{Field equations and classical solutions:}

The general field equations corresponding to action \eqref{2.1} are,

\be \label{2.2} \Big(R_{\mu\nu} - {1\over 2} g_{\mu\nu}R\Big) f(\phi) + g_{\mu\nu}\Box f(\phi)  - f_{;\mu;\nu} -{\omega(\phi)\over \phi}\phi_{,\mu}\phi_{,\nu} + {1\over 2} g_{\mu\nu}\Big(\phi_{,\alpha}\phi^{,\alpha} + V(\phi)\Big) = T_{\mu\nu},\ee
\be \label{2.3} R f' + 2 {\omega(\phi)\over \phi}\Box\phi +\Big({\omega'(\phi)\over \phi} -{\omega(\phi)\over \phi^2}\Big)\phi_{,\mu}\phi^{,\mu} - V'(\phi) = 0,\ee
where prime denotes derivative with respect to $\phi$, and $\Box$ denotes D'Alembertian, such that, $\Box f(\phi) = f''\phi_{,\mu}\phi^{^,\mu} - f'\Box\phi$. The model involves three functional parameters viz. the coupling parameter $f(\phi)$, the Brans-Dicke parameter $\omega(\phi)$ and the potential $V(\phi)$. It is customary to choose these parameters by hand in order to study the evolution of the universe. However, we have proposed a unique technique to relate the parameters in such a manner, that choosing one of these may fix the rest \cite{cons1, cons2, cons3}. We have shown that there exist a general conserved current which is admissible by the above pair of field equations, as demonstrated in \cite{Beh,Sym,cons1, cons2, cons3,cons4} leading to

\be\label{vf} V(\phi) \propto f(\phi)^2.\ee

It is convenient and hence customary to study inflationary evolution in the Einstein's frame under suitable transformation of variables, where possible. In the very early vacuum dominated era, symmetry holds, and thus we can express the action \eqref{2.1} in the form,

\be \label{3.1} A = \int \left[f(\phi) R - {K(\phi)\over 2}\phi_{,\mu}\phi^{,\mu} - V(\phi)\right]\sqrt{-g}~ d^4 x,\ee
where, $K(\phi) = 2 {\omega(\phi)\over \phi}$. Under a conformal transformation \cite{Conformal}

\be \label{confor}g_{E{\mu\nu}}= f(\phi) g_{\mu\nu},\ee
the above action \eqref{3.1} may be translated to the following Einstein's frame,

\be \label{3.2} A = \int \left[R_E - {1\over 2}\sigma_{E,\mu}{\sigma_{E}}^{,\mu} - V_E(\sigma(\phi))\right]\sqrt{-g_E}~ d^4 x,\ee
where, the subscript $`E$' stands for Einstein's frame. The effective potential $V_E$ and the transformed scalar field $\sigma$ in the Einstein's frame may be found from the following expressions,

\be \label{3.3} V_E = {V(\phi)\over f^2(\phi)};\hspace{0.4 in}\mathrm{and,}\hspace{0.4 in}\left({d\sigma\over d\phi}\right)^2 = {K(\phi)\over f(\phi)} + 3 {f'^2(\phi)\over f^2(\phi)}= {2\omega(\phi)\over \phi f(\phi)} + 3 {f'^2(\phi)\over f^2(\phi)}.\ee

\subsubsection{Inflation under Slow Roll approximation:}

Inflation with such a non-minimally coupled scalar-tensor theory of gravity is undergoing serious investigation over several decades \cite{1,2,3,4,5,6,7,8,9,10,11}. In view of the action (\ref{3.2}), one can cast the field equations, viz. the Klein-Gordon and the ($^0_0$) equations of Einstein in the background of Robertson-Walker \eqref{RW} metric as,

\be\label{FE}\begin{split}& \ddot\sigma +3\mathrm{H}\dot\sigma + k_0V_E' = 0;\hspace{0.6 in} 3\mathrm{H}^2 = \frac{1}{2} \dot\sigma^2 + k_0V_E ,\end{split}\ee
where, the Hubble parameter is defined as $\mathrm{H}={\dot a_E\over a_E}$, and $k_0 =1M^2_P$, while the slow-roll parameters and the number of e-folding take the following forms,

\be\label{3.4} \epsilon = \Big({V'_E\over V_E}\Big)^2\Big({d\sigma\over d\phi}\Big)^{-2}; \hspace{0.1 in} \eta = 2\left[\Big({V''_E\over V_E}\Big)\Big({d\sigma\over d\phi}\Big)^{-2} - \Big({V'_E\over V_E}\Big)\Big({d\sigma\over d\phi}\Big)^{-3}{d^2\sigma\over d\phi^2}\right];\hspace{0.1 in}\mathrm {N} = \int_{t_i}^{t_f} \mathrm{H} dt ={1\over 2}\int_{\phi_e}^{\phi_b} {d\phi\over \sqrt \epsilon}{d\sigma\over d\phi}.\ee
In the above, $t_i$ and $t_f$ denote time for the beginning and the end of inflation respectively.\\

\noindent
\textbf{1. Quadratic Potential:}\\

We should choose the same form of quadratic potential for the comparative study under consideration. Thus, in view of the symmetry, \eqref{vf} if we choose $f(\phi)=f_0\phi$ then $V(\phi)= m^2\phi^2+C_0$, where we have added a constant $C_0$ in the potential without loss of generality. The parameters of the theory under consideration can therefore be expressed as \cite{Sym},

\be \label{ParaA}\begin{split}& \omega(\phi) = \frac{\omega_0^2 - 3f_0^2}{2f_0},~~~{d\sigma\over d\phi}={\omega_0\over f_0\phi},\hspace{0.3 in} V_E ={1\over f_0^2} (m^2 + C_0\phi^{-2}),\hspace{0.3 in}\epsilon ={4 f_0^2C_0^2\over \omega_0^2(C_0 + m^2\phi^2)^2},\\&
\eta = {8f_0^2C_0\over \omega_0^2(C_0 + m^2\phi^2)},\hspace{0.3 in}
\mathrm{N} = \frac{\omega_0^2}{4f_0^2C_0}\left[m^2\left({\phi_i^2\over 2} - {\phi_f^2\over 2}\right) + C_0(\ln{\phi_i} - \ln{\phi_f}) \right].\end{split}\ee
In view of the above forms of the slow roll parameters \eqref{ParaA}, we present table-8, underneath, corresponding to  $m^2 > 0$. The wonderful fit with the latest data sets released by Planck \cite{Planck1,Planck2} is particularly significant because, $0.959 <  n_s < 0.970$, while $r <0.033$. Further, the number of e-fold ($40 \le \mathrm{N} \le 53$) is sufficient to alleviate the horizon and flatness problems.\\

In view of the above form of $V_E$ \eqref{3.3} and taking the values of $m^2$,~$C_0,~\phi_i$ from table-8, we obtain the following expression from equation \eqref{FE},
\be 3\mathrm{H}^2= k_0V_E = {1\over f_0^2}\left(m^2+{C_0\over \phi^2}\right)M^2_P\approx 10\times 10^{-14}M^2_P.\ee
Hence, the energy scale of inflation has been found to be sub-Planckian $\mathrm{H}* \approx 10^{-7} M_P$. Additionally, this model also gracefully exits from inflationary regime, since the scalar field exhibits oscillatory behaviour $\phi \sim  {1\over 2m^2}\left[(1-m^2C_0) \cos{\left({\sqrt {2m^2}\over \omega_0}t\right)}\right]$, at the end of inflation. Thus the scalar field starts oscillating many times over a Hubble time, driving a matter-dominated era at the end of inflation.

\begin{center}
\begin{minipage}[h]{0.47\textwidth}
      \centering
\begin{tabular}{|c|c|c|c|c|}
\hline\hline
 $\phi_f ~in M_P$  & $\omega_0 ~in M_P$ &  $r=16\epsilon$ & $n_s$ & $\mathrm{N}$ \\\hline
  1.01905 & 6.5 & 0.03192  &.9605 & 41\\\hline
 1.01802 & 6.6 & 0.03096 &.9617 & 42 \\\hline
 1.01702& 6.7 & 0.03004 &.9629 & 44 \\\hline
 1.01605& 6.8 & 0.02917 &.9639 & 45 \\\hline
 1.01510& 6.9 & 0.02833  &.9650 & 46 \\\hline
 1.01419& 7.0 & 0.02752  &.9660 & 48 \\\hline
 1.01329& 7.1 & 0.02675 &.9669 & 49 \\\hline
 1.01242& 7.2& 0.02601 &.9678 & 50 \\\hline
 1.0116& 7.3 & 0.02531 &.9687 & 52 \\\hline
 1.0107& 7.4 & 0.02463 &.9696 & 53 \\\hline
 1.0099& 7.5 & 0.02397 &.9704 & 55 \\\hline
\hline
\end{tabular}
      \captionof{table}{$f(\phi)=f_0\phi$, $f_0={1\over 2}M_P$; ${\phi_i}=2.0M_P$;\\${C_0}=-0.9\times {10^{-13}M_P^4};~{m^2}=1.0\times {10^{-13}M_P^2}.$}
      \label{table:8}
   \end{minipage}%
   \end{center}

Such astounding fit with the experimental data provokes to study the late-stage of cosmic evolution. In this connection, we mention that the choice of the quadratic form of potential was undertaken due to the fact that de-Sitter solution for all the four higher-order modified theories of gravity considered here, restricts the potential in its quadratic form only. Nonetheless, in an early work \cite{Beh}, a quartic potential was taken into account, and it was shown that it can account for the late-stage of cosmic evolution, with excellence. In the following we therefore study inflation in view of quartic potential.\\

\noindent
\textbf{2. Quartic potential:}\\

In an earlier work \cite{Beh}, relaxing the symmetry, the coupling parameter and the potential were chosen as $f (\phi)= \phi^2$, and $V(\phi)=V_0\phi^4-B\phi^2$, so that, the parameters of the theory under consideration can be expressed as,
\be \label{ParaB} \omega(\phi) = \frac{\omega_0^2 - 12\phi^2}{2\phi},~~~{d\sigma\over d\phi}={\omega_0\over \phi^2},~~~V_E = V_0 - {B\over \phi^2};\ee
\be\begin{split}& \epsilon=\frac{4B^2\phi^2}{\omega_0^2(V_0\phi^2-B)^2} ;~~~~~\eta=-\frac{4B\phi^2}{\omega_0^2(V_0\phi^2-B)};~~~~~
 \mathrm{N} = {\omega_0^2\over 4B}\left[{V_0}\ln(\phi_i - \phi_f)+{B\over 2}\left({1\over \phi_i^2} - {1\over \phi_f^2}\right)\right].\end{split}\ee
It is important to mention that the same form of potential was also considered to study late-time cosmic acceleration \cite{12}. The reason for such a choice of the potential was also clarified in \cite{Beh}. In a nut-shell: in the non-minimal theory, the flat section of the potential $V(\phi)$, responsible for slow-rollover, is usually distorted. However, generalizing the form of non-minimal coupling by an arbitrary function $f(\phi)$, Park and Yamaguchi \cite{13} could show that the flat potential required for slow roll, is still obtainable when $V_E$ is asymptotically constant. Here, initially when $\phi\gg\sqrt{B\over V_0}$, the second term may be neglected, so that $V_E \approx V_0$, and the potential becomes flat, admitting slow roll. Inflationary parameters were found to fit experimental data released (Planck's collaboration-2016) by the time the work \cite{Beh} was carried out. However, over years, Planck's data puts up tighter constraints on inflationary parameters, and so it is quite reasonable to check, if this form of the quartic potential passes the said constraint limits \cite{Planck1,Planck2}.\\

\noindent
\textbf{Case-A:}\\

In the following table 9, we present our computed results on inflationary parameters, taking ${\phi_i}=1.3M_P$, ${B}=1.0\times {10^{-20}M_P^2}$ and ${V_0}=1.1\times {10^{-20}}$. It is found that the spectral index of scalar perturbation ranges between $0.959 <  n_s < 0.970$, the tensor to scalar ratio remains around $r \approx0.06$, and the number of e-folding is around $\mathrm{N} \approx 50$, showing marvellous fit with the currently released data set \cite{Planck1,Planck2}.

\begin{center}
\begin{minipage}[h]{0.47\textwidth}
      \centering
\begin{tabular}{|c|c|c|c|c|}
 \hline\hline
 ${\phi_f}~in M_P$ &${\omega_0}~in M_P$ & $r=16\epsilon$ & $n_s$& $\mathrm{N}$ \\\hline
 .9753 &42& 0.08309 & .9599 & 38 \\\hline
 .9748 &43 &0.07927& .9618 & 40 \\\hline
 .9743 & 44 &0.07571 & .9635& 42\\\hline
 .9739 & 45 &0.07238 & .9650 & 44 \\\hline
 .9734 & 46 & 0.06927 &.9666 & 46 \\\hline
 .9730 & 47 &0.06636 & .9680 & 48 \\\hline
 .9726 & 48 & 0.06362 &.9693 & 50 \\\hline
 .9722 & 49 & 0.06105 &.9706 & 52 \\\hline
\hline
\end{tabular}
\captionof{table}{$f(\phi) = \phi^{2}$:~~~ ${\phi_i}=1.3M_P$,\\${B}=1.0\times {10^{-20}M_P^2};~{V_0}=1.1\times {10^{-20}}$.}
      \label{table:9}
   \end{minipage}
   \end{center}

Let us therefore proceed to find the energy scale of inflation. In view of the above form of $V_E$ \eqref{ParaB}, and using \eqref{3.3}, we obtain the following expression from equation \eqref{FE},

\be\label{EE} 3\mathrm{H}^2\approx {k_0}{V_E}=\left(V_0-{B\over \phi^2}\right)M^2_P \approx 50.8\times10^{-22}M^2_P.\ee
The numerical value of $3\mathrm {H}^2$ is an outcome of the values of $V_0$, $B$, and $\phi_i$ presented in table-9. Clearly, the energy scale of inflation is sub-Planckian $(\mathrm{H}* \approx 4.11\times 10^{-11} M_P)$. Exact solution of equation \eqref{FE} using Mathematica, does not evince oscillatory behaviour of the scalar field at the end of inflation. We therefore choose an oscillatory $\phi$ a-priori, and see if the consequence is physically admissible. Let us therefore assume,

\be\label{phi} \phi=\exp(iwt).\ee
Equation \eqref{FE} may therefore be expressed as,

\be 6\mathrm{H}^2=2k_0\left(V_0-{B\over \phi^2}\right)-{{w^2\omega_0^2}\over \phi^2}.\ee
Now taking  the numerical values of the parameters from table-9, viz. $V_0=1.1\times {10^{-20}},~B=1.0\times {10^{-20}}M_P^2,~\omega_0=45M_P,~\phi_f=0.973M_P$, and $k_0=1M_P$, we get,

\be 6\mathrm{H}^2\approx 8.58\times{10^{-22}}M^2_P-2139w^2M^2_P.\ee
If $ w^2= 1\times10^{-25},$ then

\be 6\mathrm{H}^2\approx 6.441\times{10^{-22}}M^2_P,~~~ \mathrm{H}^2\approx 1.07\times{10^{-22}}M^2_P.\ee
Hence, $\mathrm{H} = 1.03 \times 10^{-11} M_P$. As repeatedly mentioned, Hubble parameter remains almost constant during inflation. It is called the scale of inflation, which is $\mathrm{H}* \approx 4.11\times 10^{-11} M_P)$ in the present model. As inflation halts ($\epsilon =1$) Hubble parameter decreases fast and we observe that as it reaches one-fourth the value of the Hubble scale ($\mathrm{H} = 1.03\times 10^{-11} M_P)$), the scalar field starts oscillating many times over a Hubble time, driving a matter-dominated era at the end of inflation. Consequently, graceful exit from inflation is also evinced.\\

\noindent
\textbf{Case-B:}\\

Quartic potential under current consideration has magical enchantment. The reason is, one can change the parameters over a wide range, and yet end up with outstanding data fit. In what follows, we show that, even setting both the parameters $B$ and $V_0$ to negative values, and thereafter also interchanging their values, amazingly nice fit with Planck's data is realized. In the table 10, we present our computed results on inflationary parameters, taking ${\phi_i}=1.5M_P$, ${B}=-1.1\times {10^{-20}M_P^2}$ and ${V_0}=-1.0\times {10^{-20}}$. The spectral index of scalar perturbation is found to range between $0.960 <  n_s < 0.970$, while the tensor to scalar ratio ranges between $0.059 <  r < 0.078$, and the number of e-folding is around $\mathrm{N} \approx 50$. In table 11, on the contrary, the values of $B$ and $V_0$ are interchanged, and the data set remains almost unaltered. Clearly, the data sets in both situations exhibit magnificent fit with the currently released data set \cite{Planck1,Planck2}.
\begin{center}
\begin{minipage}[h]{0.47\textwidth}
      \centering
\begin{tabular}{|c|c|c|c|c|}
 \hline\hline
 ${\phi_f}~in M_P$ &${\omega_0}~in M_P$ & $r=16\epsilon$ & $n_s$& $\mathrm{N}$ \\\hline
 1.07598    & 41 & 0.07838 & .9604 & 38 \\\hline
 1.07533 & 42 & 0.07468 & .9622 & 41 \\\hline
 1.07470 & 43 &0.07125 & .9640& 43\\\hline
 1.07471 & 44 &0.06805 & .9656 & 45 \\\hline
 1.07354 & 45 & 0.06506 &.9671 & 47 \\\hline
 1.07299 & 46 & 0.06226 &.9685 &  49\\\hline
 1.07247 & 47 & 0.05964 &.9698 & 51 \\\hline

\hline
\end{tabular}
\captionof{table}{$f(\phi) = \phi^{2}$:~~~ ${\phi_i}=1.5M_P$,\\${B}=-1.1\times {10^{-20}M_P^2};~{V_0}=-1.0\times {10^{-20}}$.}
      \label{table:10}
   \end{minipage}%
\hfill%
   \begin{minipage}[h]{0.47\textwidth}
      \centering
\begin{tabular}{|c|c|c|c|c|}
 \hline\hline
 ${\phi_f}~in M_P$ &${\omega_0}~in M_P$ & $r=16\epsilon$ & $n_s$& $\mathrm{N}$ \\\hline
 .9753 &42& 0.08309 & .9599 & 38 \\\hline
 .9748 &43 &0.07927& .9618 & 40 \\\hline
 .9743 & 44 &0.07571 & .9635& 42\\\hline
 .9739 & 45 &0.07238 & .9650 & 44 \\\hline
 .9734 & 46 & 0.06927 &.9666 & 46 \\\hline
 .9730 & 47 &0.06636 & .9680 & 48 \\\hline
 .9726 & 48 & 0.06362 &.9693 & 50 \\\hline
 .9722 & 49 & 0.06105 &.9706 & 52 \\\hline
\hline
\end{tabular}
\captionof{table}{$f(\phi) = \phi^{2}$:~~~ ${\phi_i}=1.3M_P$,\\${B}=-1.0\times {10^{-20}M_P^2};~{V_0}=-1.1\times {10^{-20}}$.}
      \label{table:11}
   \end{minipage}
   \end{center}
The energy scale of inflation is sub-Planckian $(\mathrm{H}* \approx 10^{-11} M_P)$ taking into account $k_0=-1M_P^2$ for both the data sets presented in table 10 and table 11. The oscillatory behaviour of the scalar field is as exhibited earlier.\\

The beauty of the quartic potential with non-minimally coupled scalar-tensor theory of gravity was evinced earlier model in connection with the later stage of cosmic evolution, taking into account the thermodynamic pressure ($p$) and the energy density ($\rho$) of the baryons and the CDM \cite{Beh}. Here, we brief the outcome. In the radiation dominated era ($p = {1\over 3}\rho$), the scalar field admits a solution in the form, $\phi=\frac{\phi_0}{\sqrt{(At-t_0)}}$, while the scale factor evolves like the usual Friedmann solution, viz, $a=a_0{\sqrt{(At-t_0)}}$. In the pressure-less dust dominated era, the scale factor admits a solution $(a = a_0\mathrm{sinh}t^{2\over 3})$, which had been graphically illustrated in \cite{Beh}. The graphical representation depicts that at the early stage of the pressure-less dust $(p = 0)$ era, the universe had undergone Friedmann-like decelerated expansion $a \propto t^{2 \over 3}$, while accelerated expansion initiated at the late stage of cosmic evolution around red-shift $z \approx 0.78$, which is in perfect agreement with experimental data. Additionally, other cosmological parameters were computed and it was found that: 1. The present value of the scale factor is exactly $(a_0=1.0)$, 2. The present value of the Hubble parameter is $\mathrm{H_0}=69.24 Km. s^{-1}Mpc^{-1}$, 3. The age of the universe is ${t_0 = 13.86 Gyr}$ and hence, 4. ${\mathrm{H_0} t_0 = 1.01}$, 5. The deceleration parameter $q$ remains almost constant $q \approx 0.5$, till the value of redshift $z = 4.0$, confirming a long Friedmann-like matter dominated era, 6. The present value of deceleration parameter is $q = -0.59$, 7. The present value of the effective state parameter is therefore, $\omega_{eff0} = -0.73$, 8. Considering, as usual, that the CMBR temperature falls as $a^{-1}$, and its value at decoupling to be $T_{dec} \approx 3000 K$, the present value of it has been found to be $T_0 = 2.7255 K$. All these agree perfectly with experimental results.

\section{\bf{Concluding remarks:}}

Seeds of structure in the universe are the density variations known as the primordial fluctuations. The prevalent and most widely accepted theory that can explain the origin of the seeds of perturbation is the cosmic inflation, which occurred soon after Planck's epoch $(t_P = 10^{-43} s)$. According to inflationary paradigm, the exponential growth of the scale factor caused quantum fluctuation of the inflation field (the scalar field that we considered here) to be stretched beyond the horizon and freeze. Later, as inflation halts, these seeds of perturbation enter the horizon and form structures. Primordial fluctuations are typically described by a power spectrum, which gives the power of variation of the function of spatial scale. Both the scalar and the tensor fluctuations follow a power law. The ratio of tensor to the scalar power spectra, called the tensor to the scalar ratio, is given by $r = 2{{|\delta_h|}^2\over {|\delta_R|}^2}$, where, $|\delta_h|^2$ and $|\delta_R|^2$ are the tensor and scalar modes of perturbation respectively, and the factor $2$ arises due to the presence of two polarizations of tensor modes. While Planck's collaboration teams \cite{Planck1, Planck2} alone constrain $r < 0.1$, the combined data from other experiments viz. BAO, BICEP2, and BK15 Keck Array, tightens the constraint to $r < 0.06$. Now, the scale-dependence of the CMB power spectrum constrains the slope of the primordial scalar power spectrum, conventionally parameterized by the power-law index $n_s$, where $n_s = 1$ corresponds to a scale-invariant spectrum. The matter and baryon densities also affect the scale-dependence of the CMB spectra in a way that differs from a variation in $n_s$, leading to relatively mild degeneracies between these parameters. Assuming that the primordial power spectrum is an exact power law, we find $n_s = 0.9649\pm .0042$ which is $8\sigma$ away from scale-invariance $(n_s = 1)$. Further, BAO data also tightens the $|n_s|$ constraint by a little amount. Combining all data, viz TT, TE, EE + lowE + lensing + BK15 + BAO, $r$ is constrained even further to $r < 0.058$ with $n_s = 0.9668 \pm 0.0037$. \\

It is therefore worth to check the viability of different gravitational actions proposed over years, in connection with the currently available inflationary data sets. Note that all the experimental data are analysed with a standard model viz. the single minimally coupled scalar field model. Hence, the result is expected to vary slightly, depending on the models. In this sense, all the four higher order theories taken up in the present analysis show quite a nice fit with the experimental data sets. However, one can deselect case 3, since the energy scale of inflation is super-Planckian, although further investigation is necessary. It is also required to see if the other three models show Friedmann-like behaviour $(a\propto \sqrt t)$ in the radiation dominated era which initiated soon after the graceful exit from inflation. It is further suggestive to check if these models exhibit a long Friedmann-like pressure-less dust dominated era $(a\propto t^{2\over 3})$, after photons decoupled and prior to the recent accelerating phase. Analytical solutions are not available for these complicated models, and future task is to numerically simulate these models in the matter dominated eras. On the contrary, the non-minimally coupled scalar-tensor theory (case 5) show excellent fit with the available data and passes the tightest constraints imposed on the inflationary parameters. Since de-Sitter solution for all the higher order theories under consideration is admissible with standard square law potential $(V = m^2 \phi^2)$, we fixed it for the non-minimally coupled scalar-tensor theory of gravity too. It is therefore required to see if the model with square law potential, potentially behaves in the matter-dominated era also. Nevertheless, with a different (quartic) potential this has been achieved earlier, which showed excellent agreement with FLRW model until recently, before it enters an accelerated phase of expansion.\\

There is a recent claim for direct detection of dark energy \cite{vagnozzi}. XENON1T, operating thousands of feet underground the Italian mountain `Monte Gran Sasso', is the most sensitive detector on earth searching for WIMP (Weakly Interacting Massive Particle) dark matter. Last year, it reported 53 excess recoil electrons than estimated. This was a great puzzle. In a recent publication \cite{vagnozzi} the authors assumed interaction of dark energy with the electro-magnetic field and followed a method called chameleon screening for their analysis. They inspected the effect on the detector, if dark energy is produced in a particular region of the sun, called tachocline, where magnetic field is very strong. To their surprise they found the excess recoil electrons are the outcome of dark energy. Future experiments will be able to confirm the claim. In this sense, the non-minimally coupled scalar-tensor theory of gravity, is highly promising. However, one cannot avoid the presence of higher-order curvature invariant terms in the very early universe. It is therefore suggestive to supplement action \eqref{2.1} at least with $R^2$ term to test the outcome. This may be posed in the future.

\end{document}